\documentclass{article}

\usepackage[nonatbib,final]{neurips_2024}

\usepackage[utf8]{inputenc} 
\usepackage[T1]{fontenc}    
\usepackage{hyperref}       
\usepackage{url}            
\usepackage{booktabs}       
\usepackage{amsfonts}       
\usepackage{nicefrac}       
\usepackage{microtype}      
\usepackage{xcolor}         

\usepackage[numbers]{natbib}
\usepackage{makecell}
\usepackage{graphicx}
\usepackage[ruled,vlined,algosection]{algorithm2e}

\title{Counterfactual MRI Data Augmentation using Conditional Denoising Diffusion Generative Models}

\author{%
  Pedro Morão \\
  Instituto Superior Técnico\\
  Universidade de Lisboa\\
  Lisboa, Portugal \\
  \And
  Joao Santinha \\
  Digital Surgery LAB, Breast Unit \\ Champalimaud Foundation \\
  Faculdade de Medicina, Universidade de Lisboa \\
  Lisboa, Portugal \\
  \texttt{joao.santinha@research.fchampalimaud.org} \\
  \And 
  Yasna Forghani \\
  Digital Surgery LAB, Breast Unit \\ Champalimaud Foundation \\
  Lisboa, Portugal \\
  \And
  Nuno Loução \\
  Digital Surgery LAB, Breast Unit \\ Champalimaud Foundation \\
  Lisboa, Portugal \\
  \And
  Pedro Gouveia \\
  Digital Surgery LAB, Breast Unit \\ Champalimaud Foundation \\
  Faculdade de Medicina, Universidade de Lisboa \\
  Lisboa, Portugal \\
  \And
  Mario A. T. Figueiredo \\
  Instituto de Telecomunicações \\
  Instituto Superior Técnico \\
  Universidade de Lisboa \\
  Lisboa, Portugal \\
}

\begin{document}

\maketitle

\begin{abstract}
  Deep learning (DL) models in medical imaging face challenges in generalizability and robustness due to variations in image acquisition parameters (IAP). In this work, we introduce a novel method using conditional denoising diffusion generative models (cDDGMs) to generate counterfactual magnetic resonance (MR) images that simulate different IAP without altering patient anatomy. We demonstrate that using these counterfactual images for data augmentation can improve segmentation accuracy, particularly in out-of-distribution settings, enhancing the overall generalizability and robustness of DL models across diverse imaging conditions. Our approach shows promise in addressing domain and covariate shifts in medical imaging. The code is publicly available at \url{https://github.com/pedromorao/Counterfactual-MRI-Data-Augmentation}
\end{abstract}

\section{Introduction}

Deep learning (DL) models in medical imaging continue to face generalizability and robustness challenges. While data augmentation has been widely used to improve the performance of DL models in various fields, current augmentation techniques do not easily replicate domain, population, and covariate shifts that arise from variations in medical image scanners, acquisition settings, and patient populations. Although style transfer has been proposed as a possible solution to harmonize images across different acquisition settings and scanners, those methods usually work by mapping a source to a target domain on a pairwise basis. That approach thus leads to combinatorially growing numbers of possible combinations that exponentially increase as new scanners and acquisition protocols emerge.

Invariant based-methods, like the one proposed by \citet{arjovsky2019invariant}, offer a promising solution to mitigate performance drops under domain and covariate shifts. However, those methods often require detailed information about the environments in which the data were acquired, as well as known clinical outcomes. Advances in image generation and modification techniques could be leveraged to synthesize new images, further enforcing invariance during training.

In this study, we introduce a novel method for creating counterfactuals from existing data using conditional denoising diffusion generative models (cDDGMs). Our approach simulates the acquisition of magnetic resonance (MR) images across different scanners and image acquisition parameters (IAP). By incorporating IAP as conditioning context for the denoising diffusion generative model (DDGM), we are able to alter images without affecting the underlying patient anatomy.

We provide a detailed description of the model and methodology in Appendix A. To evaluate the effectiveness of the generated counterfactual IAP images, we use metrics such as the Fréchet inception distance (FID), structural similarity index metric (SSIM), and maximum mean discrepancy (MMD). Additionally, we assess the ability of these images to mislead a multi-task model trained to predict the IAP from MR images. Finally, we examine the impact of using these counterfactual images for data augmentation on the generalizability of DL segmentation models, focusing on both in-distribution (ID) and out-of-distribution (OOD) scenarios. 

\section{Methodology}
\subsection{Dataset}
We employed the Duke-Breast-Cancer-MRI dataset \cite{saha2018machine} to train and evaluate our deep generative model. The dataset comprised pre-contrast dynamic contrast-enhanced breast MRIs from 922 patients, with 100 patients also containing breast tissue segmentation masks. More
pre-processing and epoch details can be found in Appendix A.1.

\subsection{Image Acquisition Parameters Prediction Model}
Following the model proposed by \citet{konz2024reverse}, a ResNet-18 \cite{he2016deep} was modified to predict 7 image acquisition parameters through the final fully-connected layer. 

The four continuous ($M = 4$) IAP - Flip Angle (FA), Slice Thickness (ST), Echo Time (TE), and Repetition Time (TR) - are predicted directly using a single unit for each of them in our network's output layer. The three categorical ($K = 3$)  IAP considered - Scanner Manufacturer (SM), Field Strength (FS), and Scan Options (SO) - are converted into one-hot
encoding each with a different number of possible values/categories. For the categorical variables, with $C_k$ ($k = 1,\cdots , K$) denoting the number of categories in each categorical variable, the final layer, has a total width of $\sum_{k=1}^K C_k + M$.

The training of the IAP model involved a multi-task learning approach with the combination of loss functions for the categorical (weighted-cross-entropy losses, $\mathcal{L}_{WCE_k}$) and continuous IAP (mean squared error losses, $\mathcal{L}_{MSE}$):
\begin{equation}
    \mathcal{L} _{IAP} = \sum_{k=1}^K \mathcal{L}_{WCE_k} (\hat{y}, y) + \sum_{m=1}^M \mathcal{L}_{MSE} (\hat{y}, y).
    \label{eq:eq1}
\end{equation}

\subsection{Conditional Denoising Diffusion Generative Model}

A cDDGM was used to modify the MR images according to the IAP. The cDDGM architecture is based on the DDPM architecture \cite{ho2020denoisingdiffusionprobabilisticmodels}, which learns to reverse a Markovian diffusion process by gradually denoising an image, starting from pure-noise. 

To condition across the multiple classes, corresponding to the different IAP, we selected the classifier free-guidance (CFG) method \cite{ho2022classifierfreediffusionguidance}, as it enables controlling the strength of the alignment with the conditional contex through a guidance scale parameter and it eliminates the need for an additional classifier, as opposed to classifier guidance.

For the diffusion process,  we used $1000$ steps of the original DDPM sampler with a cosine noise scheduler.

Because we only want to modify the original images, we stop the forward diffusion process early on, at a point where the distributions of the different IAP overlap due to the perturbation that was added to the image, and then reverse the diffusion process while conditioning the image on a different set of IAP. This approach is similar to that of \cite{meng2022sdeditguidedimagesynthesis}, but with a conditional model.

\subsection{Breast Tissue Segmentation Model}

For the breast tissue segmentation, a U-Net \cite{DBLP:journals/corr/RonnebergerFB15}, using residual blocks to enable better gradient back-propagation and facilitate the optimization process, was used to segment MRI images into 3 different labels fat, fibroglandular tissue (FGT) and background.

\subsection{IAP, cDDGM, and Segmentation Models  Training}
\label{sec:training}

We used images from the 822 patients without breast segmentations to train the cDDGM and IAP models. The training of the segmentation models used the images and corresponding breast tissue segmentation masks of the remaining 100 patients, while considering different scenarios: (1) mix of images from different manufacturers available for training; (2) images from only one manufacturer available for training. More details about training are provided in Appendix A.3.

\subsection{Evaluation metrics}

Similarly to the work from \citet{konz2024reverse}, we evaluated the performance of our IAP model top-1 accuracy for categorical IAP and mean squared error for continuous IAP.

To assess the segmentation model, we used the The Dice-S{\o}rensen coefficient and accuracy for each different breast tissue present in the segmentation masks.

Given that the cDDGM was trained to perform IAP counterfactual data augmentation, for each image, we randomly sampled a set of IAP and generated new counterfactual images. In the case where the sampled set of IAP matches the original image's IAP, the model's output should preserve the image's IAP and recover the original image.

The performance of the cDDGM on this task was evaluated using the Fréchet inception distance (FID),
structural similarity metric (SSIM), maximum mean discrepancy (MMD), and the ability to "fool" the IAP prediction model and have the model predict the counterfactual IAP instead of the originally IAP.

Since the developed cDDGM was trained to perform changes in tissue contrast based on the IAP, without changing the anatomy, we then used the IAP counterfactual images as data augmentation samples and assessed the effect on the performance of the segmentation models in the two scenarios presented in section \ref{sec:training}.

\section{Results and Discussion}

Table \ref{table:table1_IAP_perf} summarizes the performance of the IAP prediction model. We see that the IAP prediction model captures with very good accuracy the IAP of the test dataset. Considering the ranges of each continuous variable (FA: [7\textdegree-12\textdegree]; ST: [1.1mm-2.5mm]; TE:[1.250ms-2.756ms]; TR: [3.540ms-7.395ms]), the IAP prediction models was able to estimate all variables with low MSE, except ST, for which the MSE was relatively higher ($\sim$ 5-12\%).

\begin{table}[ht]
\centering
\caption{Model prediction performance for all IAP on the Test Set. An upward arrow indicates that a higher value is better, and vice versa.}
\label{table:table1_IAP_perf}
\begin{tabular}{lcc}
\Xhline{3\arrayrulewidth}
\begin{tabular}[c]{@{}l@{}}Image acquisition\\ parameter (IAP)\end{tabular} & \multicolumn{1}{l}{\begin{tabular}[c]{@{}l@{}}Top-1 pred. \\ acc. (\%) $\uparrow$ \end{tabular}} & \multicolumn{1}{l}{Pred. MSE $\downarrow$ } \\ \Xhline{2\arrayrulewidth}
Manufacturer Model                                                          & 98.9                                                                                 & NA                            \\ \hline
Field Strength                                                              & 99.2                                                                                 & NA                            \\ \hline
Scan Options                                                                & 99.9                                                                                 & NA                            \\ \hline
Flip Angle (º)                                                              & NA                                                                                   & 0.080                         \\ \hline
Slice Thickness (mm)                                           & NA                                                                                   & 0.133                         \\ \hline
TE (ms)                                                                     & NA                                                                                   & 0.005                         \\ \hline
TR (ms)                                                                     & NA                                                                                   & 0.046                         \\ \Xhline{3\arrayrulewidth}
\end{tabular}
\end{table}

Table \ref{table:table2_Segmentations} presents the segmentation accuracies for background, fat, and FGT, along with the mean Dice scores for models trained using images from GE and Siemens MRI scanners. Additionally, it includes results for in-distribution (ID) settings (e.g., trained on GE, applied to GE; trained on Siemens, applied to Siemens) and out-of-distribution (OOD) settings (e.g., trained on GE, applied to Siemens; trained on Siemens, applied to GE). These results are based on a model selected from various configurations of number of steps and guidance scales, which are detailed in Tables \ref{table:table3_Gen_Evaluation_perf} and \ref{table:table4_GenIAP_perf} in Appendix \ref{sec:appendix}.

\begin{table}[ht]
\centering
\caption{Segmentation performance in ID and OOD cases with and without counterfactual IAP data augmentation.}
\label{table:table2_Segmentations}
\begin{tabular}{lcccccc} \Xhline{3\arrayrulewidth}
Setting                                                                             & \multicolumn{1}{l}{\begin{tabular}[c]{@{}l@{}}Acc.\\ Background $\uparrow$\end{tabular}} & \multicolumn{1}{l}{\begin{tabular}[c]{@{}l@{}}Acc.\\Fat $\uparrow$\end{tabular}} & \multicolumn{1}{l}{\begin{tabular}[c]{@{}l@{}}Acc.\\FGT $\uparrow$\end{tabular}} & \multicolumn{1}{l}{\begin{tabular}[c]{@{}l@{}}Dice\\Fat $\uparrow$\end{tabular}} & \multicolumn{1}{l}{\begin{tabular}[c]{@{}l@{}}Dice\\FGT $\uparrow$\end{tabular}} & \multicolumn{1}{l}{\begin{tabular}[c]{@{}l@{}}Mean\\Dice $\uparrow$\end{tabular}} \\ \Xhline{3\arrayrulewidth}
Both                                                                                & 99.4                                                                          & 93.6                                                                   & 90.3                                                                   & 0.929                                                                  & 0.716                                                                  & 0.840                                                                   \\ \Xhline{2\arrayrulewidth}
GE trained in GE                                                                    & 99.4                                                                          & \textbf{96.6}                                                                   & 88.6                                                                   & 0.949                                                                  & \textbf{0.758}                                                         & \textbf{0.863}                                                          \\ 
\begin{tabular}[c]{r}  cDDGM - \# steps:50; gs:3\end{tabular}           & \textbf{99.5}                                                                 & 96.5                                                                   & \textbf{89.4}                                                          & \textbf{0.950}                                                         & 0.757                                                                  & \textbf{0.863}                                                          \\ \Xhline{2\arrayrulewidth}
SIEMENS trained in GE                                                               & 98.3                                                                          & 94.0                                                                   & \textbf{62.3}                                                          & 0.860                                                                  & \textbf{0.555}                                                         & 0.730                                                                   \\ 
\begin{tabular}[c]{r}  cDDGM - \# steps:50; gs:3\end{tabular}      & \textbf{99.1}                                                                 & \textbf{94.2}                                                                   & 61.7                                                                   & \textbf{0.889}                                                         & 0.536                                                                  & \textbf{0.739}                                                          \\ \Xhline{2\arrayrulewidth}
SIEMENS trained in SIEMENS                                                          & \textbf{99.4}                                                                 & 91.5                                                                   & 58.7                                                                   & \textbf{0.866}                                                         & \textbf{0.588}                                                         & \textbf{0.746}                                                          \\ 
\begin{tabular}[c]{r}  cDDGM - \# steps:50; gs:3\end{tabular} & 99.1                                                                          & \textbf{92.3}                                                          & \textbf{60.2}                                                                   & 0.863                                                                  & 0.571                                                                  & 0.737                                                                   \\ \Xhline{2\arrayrulewidth}
GE trained in SIEMENS                                                               & \textbf{98.9}                                                                 & 89.3                                                                   & 59.7                                                                   & 0.886                                                                  & 0.549                                                                  & 0.742                                                                   \\ 
\begin{tabular}[c]{r}  cDDGM - \# steps:50; gs:3\end{tabular}      & 98.8                                                                          & \textbf{91.8}                                                          & \textbf{67.7}                                                                   & \textbf{0.896}                                                         & \textbf{0.553}                                                                  & \textbf{0.750}                                                                 \\ \Xhline{3\arrayrulewidth}
\end{tabular}
\end{table}

The results indicate that the use of IAP counterfactual images leads to slight improvements in accuracy for background, FGT, and the Dice score for fat in an ID setting with GE scanners. A similar improvement was observed for fat and FGT with Siemens scanners. In OOD settings, when the model was trained with GE images, the inclusion of IAP counterfactual images positively impacted the accuracies for background and fat, as well as the mean and fat Dice scores. In the OOD setting, when the model was trained on Siemens images and applied to GE images, the IAP counterfactual model improved the segmentation accuracy for fat and FGT, along with enhancing the fat, FGT, and mean Dice scores.

\begin{figure}[ht]
  \centering
  \caption{Comparison between the ground truth (True) and DL breast segmentation models trained without data augmentation (Pred.) and with data augmentation using cDDGM (Pred.Aug.), in out-of-distribution settings. (A) results of models trained in GE when applied to Siemens MRIs. (B) results of models trained in Siemens when applied to GE MRIs. Blue - Fat mask; Orange - FGT mask.}
  \includegraphics[width=0.45\textwidth]{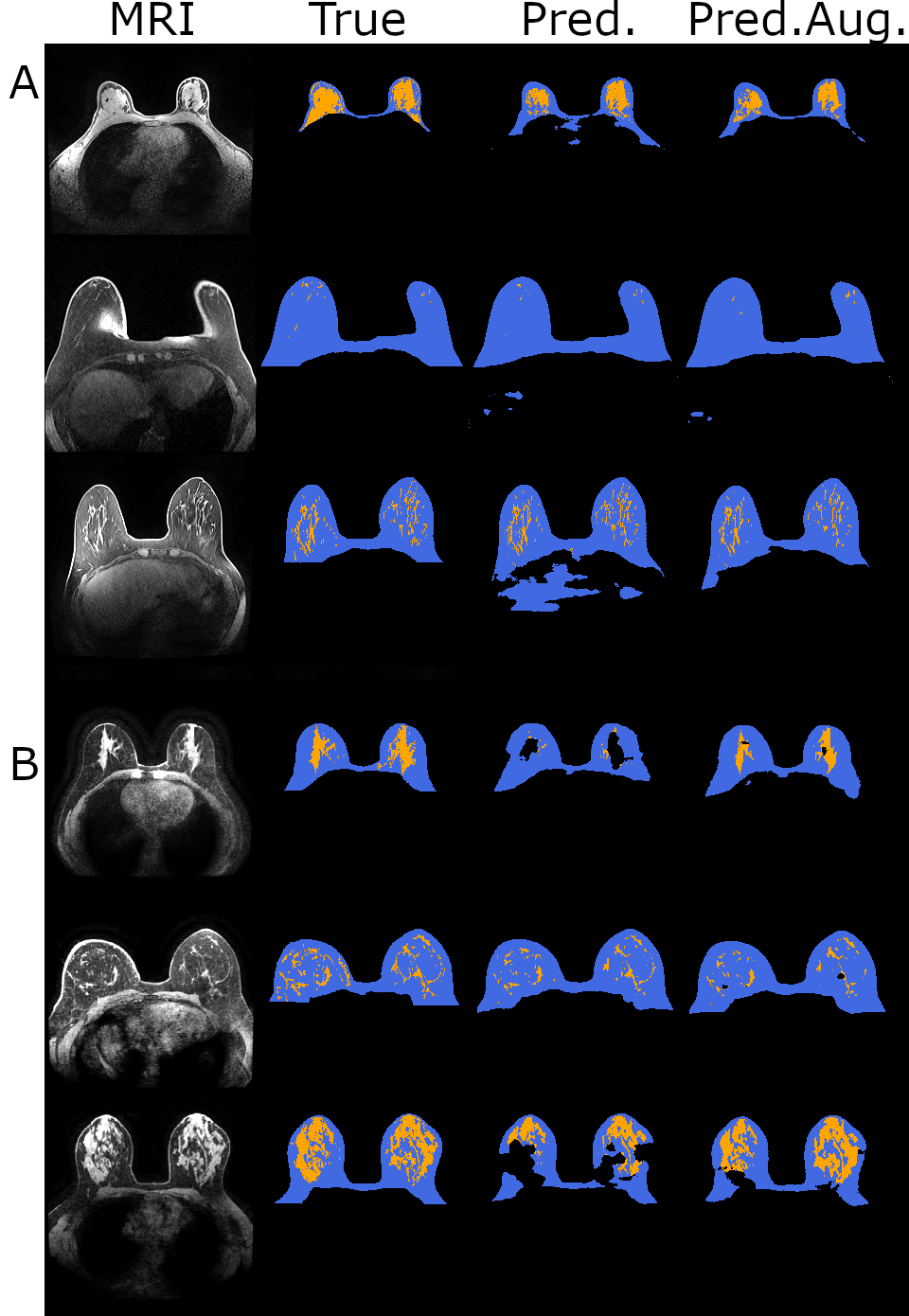}
  \label{fig:figure_1_seg_ood}
\end{figure}

Figure \ref{fig:figure_1_seg_ood} showcases several examples of breast MRIs from the test set, along with corresponding ground truth tissue masks and DL segmentation predictions without and with IAP counterfactual images used as data augmentation in the two OOD settings previously mentioned. In Figure 1-A, we observe that the DL segmentation model without cDDGM data augmentation has a propensity to incorrectly classify background areas (black) in the chest wall (top image) and liver (third image), where the model with cDDGM data augmentation was able to reduce these errors. As for Figure 1-B, we see several holes in the breast tissue masks of the first and third predictions using the DL segmentation model trained without cDDGM data augmentation that are reduced when the proposed data augmentation method is used. Figure 2 in the Appendix \ref{sec:appendix} demonstrates that well in both ID scenarios. 

Our work is limited by the lack of diversity in MRI scanner manufacturers and the dataset size (e.g., of the 100 patients containing breast tissue segmentations, only 29 patients were acquired in Siemens scanners). Nevertheless the use of the proposed cDDGM for counterfactual MRI data augmentation yielded promising results, demonstrating its potential to improve generalizability and robustness of DL models in medical imaging.

\section{Conclusions}

In this work, we demonstrated that integrating IAP counterfactual images using cDDGM can enhance the generalizability and robustness of deep learning models in medical imaging. The generated counterfactual images successfully misled the IAP prediction model into predicting the intended counterfactual parameters. Moreover, using these images for data augmentation led to slight improvements in segmentation accuracy, particularly in out-of-distribution (OOD) settings, thereby improving the generalizability of DL models across diverse medical imaging conditions.

\bibliographystyle{unsrtnat}
\bibliography{neurips_2024}

\begin{thebibliography}{13}
\providecommand{\natexlab}[1]{#1}
\providecommand{\url}[1]{\texttt{#1}}
\expandafter\ifx\csname urlstyle\endcsname\relax
  \providecommand{\doi}[1]{doi: #1}\else
  \providecommand{\doi}{doi: \begingroup \urlstyle{rm}\Url}\fi

\bibitem[Arjovsky et~al.(2019)Arjovsky, Bottou, Gulrajani, and Lopez-Paz]{arjovsky2019invariant}
Martin Arjovsky, L{\'e}on Bottou, Ishaan Gulrajani, and David Lopez-Paz.
\newblock Invariant risk minimization.
\newblock \emph{arXiv preprint arXiv:1907.02893}, 2019.

\bibitem[Saha et~al.(2018)Saha, Harowicz, Grimm, Kim, Ghate, Walsh, and Mazurowski]{saha2018machine}
Ashirbani Saha, Michael~R Harowicz, Lars~J Grimm, Connie~E Kim, Sujata~V Ghate, Ruth Walsh, and Maciej~A Mazurowski.
\newblock A machine learning approach to radiogenomics of breast cancer: a study of 922 subjects and 529 dce-mri features.
\newblock \emph{British journal of cancer}, 119\penalty0 (4):\penalty0 508--516, 2018.

\bibitem[Konz and Mazurowski(2024)]{konz2024reverse}
Nicholas Konz and Maciej~A Mazurowski.
\newblock Reverse engineering breast mris: Predicting acquisition parameters directly from images.
\newblock In \emph{Medical Imaging with Deep Learning}, pages 829--845. PMLR, 2024.

\bibitem[He et~al.(2016)He, Zhang, Ren, and Sun]{he2016deep}
Kaiming He, Xiangyu Zhang, Shaoqing Ren, and Jian Sun.
\newblock Deep residual learning for image recognition.
\newblock In \emph{Proceedings of the IEEE conference on computer vision and pattern recognition}, pages 770--778, 2016.

\bibitem[Ho et~al.(2020)Ho, Jain, and Abbeel]{ho2020denoisingdiffusionprobabilisticmodels}
Jonathan Ho, Ajay Jain, and Pieter Abbeel.
\newblock Denoising diffusion probabilistic models, 2020.
\newblock URL \url{https://arxiv.org/abs/2006.11239}.

\bibitem[Ho and Salimans(2022)]{ho2022classifierfreediffusionguidance}
Jonathan Ho and Tim Salimans.
\newblock Classifier-free diffusion guidance, 2022.
\newblock URL \url{https://arxiv.org/abs/2207.12598}.

\bibitem[Meng et~al.(2022)Meng, He, Song, Song, Wu, Zhu, and Ermon]{meng2022sdeditguidedimagesynthesis}
Chenlin Meng, Yutong He, Yang Song, Jiaming Song, Jiajun Wu, Jun-Yan Zhu, and Stefano Ermon.
\newblock Sdedit: Guided image synthesis and editing with stochastic differential equations, 2022.
\newblock URL \url{https://arxiv.org/abs/2108.01073}.

\bibitem[Ronneberger et~al.(2015)Ronneberger, Fischer, and Brox]{DBLP:journals/corr/RonnebergerFB15}
Olaf Ronneberger, Philipp Fischer, and Thomas Brox.
\newblock U-net: Convolutional networks for biomedical image segmentation.
\newblock \emph{CoRR}, abs/1505.04597, 2015.

\bibitem[Rombach et~al.(2022)Rombach, Blattmann, Lorenz, Esser, and Ommer]{rombach2022highresolutionimagesynthesislatent}
Robin Rombach, Andreas Blattmann, Dominik Lorenz, Patrick Esser, and Björn Ommer.
\newblock High-resolution image synthesis with latent diffusion models, 2022.
\newblock URL \url{https://arxiv.org/abs/2112.10752}.

\bibitem[Pinaya et~al.(2022)Pinaya, Tudosiu, Dafflon, da~Costa, Fernandez, Nachev, Ourselin, and Cardoso]{pinaya2022brainimaginggenerationlatent}
Walter H.~L. Pinaya, Petru-Daniel Tudosiu, Jessica Dafflon, Pedro~F da~Costa, Virginia Fernandez, Parashkev Nachev, Sebastien Ourselin, and M.~Jorge Cardoso.
\newblock Brain imaging generation with latent diffusion models, 2022.
\newblock URL \url{https://arxiv.org/abs/2209.07162}.

\bibitem[Paszke et~al.(2019)Paszke, Gross, Massa, Lerer, Bradbury, Chanan, Killeen, Lin, Gimelshein, Antiga, et~al.]{paszke2019pytorch}
Adam Paszke, Sam Gross, Francisco Massa, Adam Lerer, James Bradbury, Gregory Chanan, Trevor Killeen, Zeming Lin, Natalia Gimelshein, Luca Antiga, et~al.
\newblock Pytorch: An imperative style, high-performance deep learning library.
\newblock \emph{Advances in neural information processing systems}, 32, 2019.

\bibitem[Cardoso et~al.(2022)Cardoso, Li, Brown, Ma, Kerfoot, Wang, Murrey, Myronenko, Zhao, Yang, et~al.]{cardoso2022monai}
M~Jorge Cardoso, Wenqi Li, Richard Brown, Nic Ma, Eric Kerfoot, Yiheng Wang, Benjamin Murrey, Andriy Myronenko, Can Zhao, Dong Yang, et~al.
\newblock Monai: An open-source framework for deep learning in healthcare.
\newblock \emph{arXiv preprint arXiv:2211.02701}, 2022.

\bibitem[Pinaya et~al.(2023)Pinaya, Graham, Kerfoot, Tudosiu, Dafflon, Fernandez, Sanchez, Wolleb, Da~Costa, Patel, et~al.]{pinaya2023generative}
Walter~HL Pinaya, Mark~S Graham, Eric Kerfoot, Petru-Daniel Tudosiu, Jessica Dafflon, Virginia Fernandez, Pedro Sanchez, Julia Wolleb, Pedro~F Da~Costa, Ashay Patel, et~al.
\newblock Generative ai for medical imaging: extending the monai framework.
\newblock \emph{arXiv preprint arXiv:2307.15208}, 2023.

\end{thebibliography}

\newpage


\appendix

\section{Appendix}
\label{sec:appendix}

\subsection{Data Normalization and Preprocessing}

The Duke-Breast-Cancer-MRI dataset comprises multiple 3D and 4D MRI sequences. Since each sequence is associated with only one set of IAP, pairwise supervised image modification techniques are not applicable. Following the approach of \citet{konz2024reverse}, we focused on the 3D pre-contrast phase of 4D dynamic contrast-enhanced sequences. In 100 patients, this phase included corresponding 3D fat and fibroglandular tissue segmentations, enabling us to evaluate the impact of the proposed cDDGM model as a data augmentation technique for DL segmentation models.

Although the selected phase represents a 3D volume, due to the size of the cDDGM model and hardware constraints, we developed and evaluated our model using 2D slices extracted from the 3D volumes. The first and last 20 slices were discarded, as they typically contained more noise and lacked relevant information.

We performed image normalization by resizing the images to 224x224 to ensure a fixed size and to accelerate model training and optimization. Although more complex models, such as Latent Diffusion Models, could handle larger images, they would require additional training of encoder and decoder networks to obtain smaller latent space in which the diffusion process would be executed. Moreover, the encoder and decoder would also need to preserve IAP-related information to ensure that the latent representation would still contain such information.

Image intensity values were normalized using percentile normalization, setting the 10th percentile to 0 and the 99th percentile to 1, without clipping values. The lower percentile was adjusted to a higher value due to the large number of low-intensity voxels in the background and thoracic cavity, which are not particularly relevant for the cDDGM or breast tissue segmentation model.

To normalize the IAP, categorical features were one-hot encoded, and numeric features were normalized by dividing by the maximum value in the dataset. This approach was chosen over min-max normalization to create a gap from 0 to the ratio of $value_{min} / value_{max}$, allowing the model to use 0 as the unconditional value. And since $value_{min} \not = 0$, this is always achievable.

\subsection{Train, Validation and Test sets}

We initially divided the dataset into subsets of images of patients with segmentations and without. We used the subset without segmentations for training and evaluating the IAP prediction model and the cDDGM model, while the subset with segmentations was exclusively used for training and evaluating the segmentation models. An iterative method was used to split the images into training, validation, and test sets, ensuring that different combinations of IAP were equally represented across all sets. Additionally, the training/validation/test splitting procedure ensured that images from the same patient were not included in different sets.

For the subset without segmentations, the dataset was split into 75\% for training, 10\% for validation, and 15\% for testing. In the subset with segmentations, 75\% of the data was used for training and 25\% used for validation. Due to the limited number of patients with segmentations, the validation set was also used as the test set to evaluate the segmentation model in an ID setting. All OOD images were used as the test set in the OOD evaluation, as they were not included in the training process.

\subsection{cDDGM Arquitecture}
\label{sec:cddgm_arch}

The noise estimation model used during the reverse diffusion process is a conditional UNet. Its architecture is inspired by the UNet design from Latent Diffusion Models \cite{rombach2022highresolutionimagesynthesislatent}. This UNet architecture incorporates cross-attention mechanisms, which enhance the model's ability to condition on complex contexts. Without these attention mechanisms, the model would struggle to effectively handle conditioning contexts that are more complex than simple image classes.

Our UNet architecture consists of six downsampling levels, one middle level, and six upsampling levels, with each level containing two residual convolution blocks. Cross-attention blocks are included on the third and fifth of the downsampling levels, on the middle level, and the corresponding positions in the upsampling levels. The conditioning is performed by adding the IAP embedding to the time embeddings and incorporating it through the cross-attention blocks. While adding more cross-attention blocks improved the model's performance, it also significantly increased the computational resources, particularly when added to earlier levels of the UNet. This 'hybrid' conditioning approach, which combines adding the condition embedding to the time embeddings and cross-attention blocks, is similar to the method used in \cite{pinaya2022brainimaginggenerationlatent}.

The model was trained using the simplified loss function from \cite{ho2020denoisingdiffusionprobabilisticmodels}, as shown in equation \ref{eqq:training_loss_cDDPM}. This loss function was adapted for the conditional training scenario, allowing the model $\mathbf{\epsilon}_{\theta}$ to receive the IAP conditioning as input but still work in an unconditional setting without $c$.
\begin{equation}
    \mathcal{L}(\theta) := \mathbb{E}_{t,\mathbf{x}_0,\mathbf{\epsilon} \sim N(0, \mathbf{I})}  \left[ \left\| \epsilon - \mathbf{\epsilon}_{\theta} (\sqrt{\bar{\alpha_t}}\mathbf{x}_0 + \sqrt{1-\bar{\alpha_t}} \mathbf{\epsilon}, c, t)  \right\|^2    \right]
\label{eqq:training_loss_cDDPM}
\end{equation}

The training algorithm for the cDDGM is equal to the original DDPM training algorithm \cite{ho2020denoisingdiffusionprobabilisticmodels} except the model is conditioned on the IAP with a conditional dropout of 15\%. The algorithm to counterfactually modify images and simulate their acquisition with other IAP is shown in \ref{alg:image_mod_alg}. Initially, noise is added to the original image $\mathbf{x}_0$ until we reach $t=steps$, then we use the CFG method \cite{ho2022classifierfreediffusionguidance} to denoise the image from $t=steps$ back to $t=0$, now conditioning the image on a new set of IAP, $c_{new}$, and controlling the guidance scale with a parameter $w$. After denoising $\mathbf{x}_{steps}$, we return the modified $\mathbf{x}_0$ with its IAP changed.

\begin{algorithm}
\caption{IAP modification algorithm using CFG}\label{alg:image_mod_alg}
$\mathbf{z} \sim \mathcal{
N}(0,\mathbf{I})$ \\
$\mathbf{x}_{steps} = \sqrt{\bar{\alpha}_{steps}}\mathbf{x}_0 + \sqrt{1-\bar{\alpha}_{steps}} \mathbf{z}$ \\

\For{$t=steps,\cdots,0$}{
$\mathbf{z} \sim \mathcal{N}(0,\mathbf{I})$ if $t > 0$, else $\mathbf{z} = 0$ \\ 
$\tilde{\mathbf{\epsilon}_t} = (1-w)\mathbf{\epsilon}_{\theta}(\mathbf{x}_t,t) + w\mathbf{\epsilon}_{\theta}(\mathbf{x}_t,c_{new},t)$\\

$\mathbf{x}_{t-1} = \frac{1}{\sqrt{\alpha_t}}(\mathbf{x}_t - \frac{1-\alpha_t}{\sqrt{1- \bar{\alpha_t}}}\tilde{\mathbf{\epsilon}_t}) + \sigma_t \mathbf{z}$}
\Return{$\mathbf{x}_0$}
\end{algorithm}

\subsection{Training Setup}

The training of all deep learning models was carried out using PYTORCH \cite{paszke2019pytorch}, MONAI CORE \cite{cardoso2022monai},
and MONAI GENERATIVE \cite{pinaya2023generative} libraries.

The training processes were conducted on a single NVIDIA A6000 GPU with 48 GB of memory.

The IAP prediction model was trained using a batch size of 512 over 200 epochs. The Adam optimizer was employed with a learning rate of $10^{-5}$ and a weight decay parameter of $10^{-4}$. The training and testing phases combined took approximately 5 hours.

For the cDDGM model, a batch size of 32 was used, with training spanning 15 epochs. The Adam optimizer was again utilized this time with a learning rate of $10^{-4}$ and weight decay of $10^{-3}$. The training took 9 hours and the testing the IAP modifications applied to the test set took from 2 to 7 hours for configuration of steps and guidance scales, varying with the number of steps. The number of channels per layer can be seen in table \ref{tab:channels_down_levels}. The cDDGM also had 8 cross-attention heads.

\begin{table}[h]
\centering

\begin{tabular}{lllllll}
\hline
cDDGM             & 64 & 64 & 128 & 128 & 256 & 256 \\ \hline
Segmentation UNet & 32 & 64 & 128 & 256 & 512 & 512 \\ \hline
\end{tabular}
\caption{Number of channels present in each down-sampling level. (the up-sampling number of channels are the same but in reverse order).}
\label{tab:channels_down_levels}
\end{table}

Data augmentation was performed, with processing times ranging from 3 to 10 hours, depending on the manufacturer and the number of steps specified. The segmentation model training, which followed the data augmentation, typically took from 30 minutes up to 1 hour, with the models trained on the larger GE subset requiring more time. The Adam optimizer with a learning rate of 0.002, a weight decay of 0.001, and a batch size of 256 were used to train the segmentation models. Early stopping was applied to determine the optimal stopping point during training. The details about the segmentation model number of channels per level can also be seen in table \ref{tab:channels_down_levels}.

\subsection{cDDGM Optimization}

The guidance scale and number of steps of the proposed cDDGM were optimized for counterfactual IAP modification. The performance of the model with the different hyperparameters was assessed using generative and similarity metrics, shown in Table \ref{table:table3_Gen_Evaluation_perf}, along with the IAP prediction model performance, shown in Table \ref{table:table4_GenIAP_perf}.

\begin{table}[ht]
\centering
\caption{cDDGM performance metrics on the IAP modification. FID: Fréchet inception distance, SSIM: Structural similarity index metric, MMD: Maximum mean discrepancy. SSIM$_{orig.\ and\ mod.}$ represents structural similarity index between original images and corresponding modified images. SSIM$_{shuff.\ and\ mod.}$ represents the structural similarity index between images from which the IAP were originally obtained and the images modified by our cDDGM using those IAP as conditioning - importantly, the images being compared were not from the same patients. An upward arrow indicates that a higher value is better, and vice versa.}
\label{table:table3_Gen_Evaluation_perf}
\begin{tabular}{lcccc}
\hline
Hyperparameters     & FID $\downarrow$  & \begin{tabular}[c]{@{}c@{}}SSIM$_{orig.\ and\ mod.}$$\uparrow$\end{tabular} & \begin{tabular}[c]{@{}c@{}}SSIM$_{shuff.\ and\ mod.}$$\downarrow$\end{tabular} & MMD $\downarrow$            \\ \hline
Without cDPPM       & 0     & 1.000                                                               & 0.258                                                               & 0               \\ \hline
\# steps: 25; gs: 3 & 0.416 & 0.742                                                               & 0.284                                                               & 0.010x10$^{-3}$ \\ \hline
\# steps: 25; gs: 5 & 0.501 & 0.709                                                               & 0.277                                                               & 0.016x10$^{-3}$ \\ \hline
\# steps: 25; gs: 7 & 0.590 & 0.689                                                               & 0.272                                                               & 0.022x10$^{-3}$ \\ \hline
\# steps: 50; gs: 3 & 0.513 & 0.657                                                               & 0.287                                                               & 0.016x10$^{-3}$ \\ \hline
\# steps: 50; gs: 5 & 0.606 & 0.630                                                               & 0.281                                                               & 0.025x10$^{-3}$ \\ \hline
\# steps: 50; gs: 7 & 0.702 & 0.613                                                               & 0.276                                                               & 0.037x10$^{-3}$ \\ \hline
\# steps: 75; gs: 3 & 0.573 & 0.606                                                               & 0.288                                                               & 0.029x10$^{-3}$ \\ \hline
\# steps: 75; gs: 5 & 0.669 & 0.583                                                               & 0.283                                                               & 0.036x10$^{-3}$ \\ \hline
\# steps: 75; gs: 7 & 0.774 & 0.566                                                               & 0.279                                                               & 0.049x10$^{-3}$ \\ \hline
\end{tabular}
\end{table}

In the table \ref{table:table3_Gen_Evaluation_perf} and \ref{table:table4_GenIAP_perf}, the row 'Without cDDGM' represents the baseline case where the model is not being applied to the images so the IAP are just being shuffled randomly for the computation of SSIM$_{shuff.\ and\ mod.}$ and the prediction of the IAP. 

\begin{table}[ht]
\centering
\caption{Model Prediction Performance for all IAP on the Test Set. An upward arrow indicates that a higher value is better, and vice versa.}
\label{table:table4_GenIAP_perf}
\begin{tabular}{lccccccc}
\hline
Hyperparam.                                                  & \multicolumn{1}{l}{\begin{tabular}[c]{@{}l@{}}MM Top-1\\ pred. acc.\\ (\%) $\uparrow$\end{tabular}} & \multicolumn{1}{l}{\begin{tabular}[c]{@{}l@{}}FS Top-1 \\ pred. acc.\\ (\%) $\uparrow$\end{tabular}} & \multicolumn{1}{l}{\begin{tabular}[c]{@{}l@{}}SO Top-1 \\ pred. acc.\\ (\%) $\uparrow$\end{tabular}} & \multicolumn{1}{l}{\begin{tabular}[c]{@{}l@{}}FA \\ Pred.\\ MSE $\downarrow$\end{tabular}} & \multicolumn{1}{l}{\begin{tabular}[c]{@{}l@{}}ST \\ Pred.\\ MSE $\downarrow$\end{tabular}} & \multicolumn{1}{l}{\begin{tabular}[c]{@{}l@{}}TR\\ Pred.\\ MSE $\downarrow$\end{tabular}} & \multicolumn{1}{l}{\begin{tabular}[c]{@{}l@{}}TE \\ Pred.\\ MSE $\downarrow$\end{tabular}} \\ \hline
\begin{tabular}[c]{@{}l@{}}Without\\ cDPPM\end{tabular}      & 22.8                                                                                     & 49.8                                                                                      & 28.0                                                                                      & 0.440                                                                         & 0.260                                                                         & 1.046                                                                        & 0.481                                                                         \\ \hline
\begin{tabular}[c]{@{}l@{}}\# steps: 25\\ gs: 3\end{tabular} & 77.7                                                                                     & 66.3                                                                                      & 87.8                                                                                      & 0.289                                                                         & 0.198                                                                         & 0.293                                                                        & 0.070                                                                         \\ \hline
\begin{tabular}[c]{@{}l@{}}\# steps: 25\\ gs: 5\end{tabular} & 82.4                                                                                     & 71.7                                                                                      & 92.8                                                                                      & 0.292                                                                         & 0.192                                                                         & 0.249                                                                        & 0.055                                                                         \\ \hline
\begin{tabular}[c]{@{}l@{}}\# steps: 25\\ gs: 7\end{tabular} & 83.4                                                                                     & 76.0                                                                                      & 93.0                                                                                      & 0.289                                                                         & 0.189                                                                         & 0.237                                                                        & 0.058                                                                         \\ \hline
\begin{tabular}[c]{@{}l@{}}\# steps: 50\\ gs: 3\end{tabular} & 85.8                                                                                     & 76.9                                                                                      & 96.0                                                                                      & 0.282                                                                         & 0.190                                                                         & 0.207                                                                        & 0.038                                                                         \\ \hline
\begin{tabular}[c]{@{}l@{}}\# steps: 50\\ gs: 5\end{tabular} & 87.2                                                                                     & 83.1                                                                                      & 96.3                                                                                      & 0.284                                                                         & 0.185                                                                         & 0.186                                                                        & 0.040                                                                         \\ \hline
\begin{tabular}[c]{@{}l@{}}\# steps: 50\\ gs: 7\end{tabular} & 88.3                                                                                     & 87.0                                                                                      & 96.6                                                                                      & 0.275                                                                         & 0.180                                                                         & 0.179                                                                        & 0.042                                                                         \\ \hline
\begin{tabular}[c]{@{}l@{}}\# steps: 75\\ gs: 3\end{tabular} & 87.1                                                                                     & 82.8                                                                                      & 97.0                                                                                      & 0.285                                                                         & 0.189                                                                         & 0.183                                                                        & 0.034                                                                         \\ \hline
\begin{tabular}[c]{@{}l@{}}\# steps: 75\\ gs: 5\end{tabular} & 88.5                                                                                     & 88.5                                                                                      & 96.5                                                                                      & 0.273                                                                         & 0.181                                                                         & 0.174                                                                        & 0.039                                                                         \\ \hline
\begin{tabular}[c]{@{}l@{}}\# steps: 75\\ gs: 7\end{tabular} & 89.3                                                                                     & 90.4                                                                                      & 96.6                                                                                      & 0.257                                                                         & 0.172                                                                         & 0.172                                                                        & 0.042                                                                         \\ \hline
\end{tabular}
\end{table}

Table \ref{table:table4_GenIAP_perf} shows that increasing the guidance scale and the amount of noise - through additional forward diffusion steps applied to the original image - enhances the proposed model's ability to predict the IAP used to generate counterfactual images. However, this comes at the cost of reduced image quality, as indicated by higher FID scores, lower SSIM$_{orig.\ and\ mod.}$ values, and increased MMD, as observed in Table \ref{table:table3_Gen_Evaluation_perf}. Specifically, the decline in SSIM between the original and modified images suggest that higher guidance scales and more diffusion steps lead to greater loss of the original image anatomical structure. Despite these changes, the MMD remains very small, indicating that the modified images stay close to the desired distribution after IAP modification. Additionally, the SSIM between the images of different patients from which the IAP were extrated to condition the image modification ($shuffled$) and the modified images remains low, suggesting that the proposed cDDGM is not altering the structure to resemble that of the $shuffled$ reference image.

\begin{figure}[ht]
  \centering
  \caption{Comparison between the ground truth (True) and DL breast segmentation models trained without data augmentation (Pred.) and with data augmentation using cDDGM (Pred.Aug.), in in-distribution settings. (A) results of models trained in GE when applied to GE MRIs. (B) results of models trained in GE when applied to GE MRIs. Blue - Fat mask; Orange - FGT mask.}
  \includegraphics[width=0.5\textwidth]{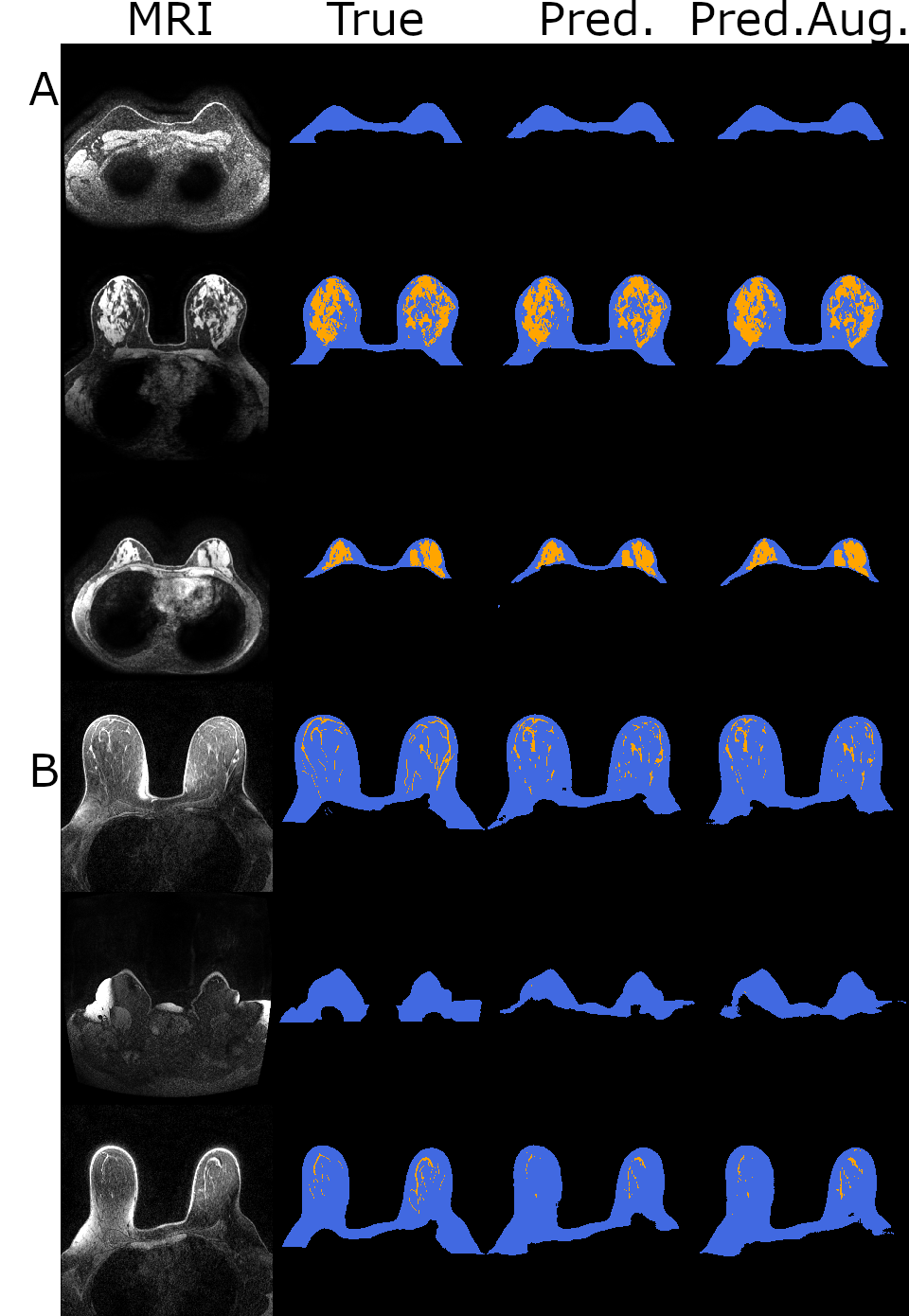}
  \label{fig:figure_2_seg_id}
\end{figure}

Figure \ref{fig:figure_2_seg_id} shows several examples of breast MRIs from the test sets, along with corresponding ground truth tissue masks and segmentation predictions without and with IAP counterfactual images used as data augmentation in the two ID settings. In both scenarios, A and B, corresponding to training and inference on images from GE and Siemens, respectively, we can observe that the DL segmentation models with and without cDDGM data augmentation perform similarly and are able approximate the ground truth.

\subsection{Data, Models' Weights and Code}

Derived data, obtained from the Duke-Breast-Cancer-MRI dataset \cite{saha2018machine}, and models' weights are made available at \url{https://zenodo.org/records/13495922}. Code is available at \url{https://github.com/pedromorao/Counterfactual-MRI-Data-Augmentation}.





\newpage

\end{document}